\documentclass[aps,prl,twocolumn,showpacs,groupedaddress,amsmath,amssymb]{revtex4}
\usepackage{graphicx}% Include figure files
\usepackage{color}
\usepackage{bm}
\usepackage{doi}
\usepackage{float}
\usepackage{mathtools}
\usepackage{nccmath}
\usepackage{physics}

\newcommand{\rd}[1]{\color{red} #1}

\begin{document}

\title{Observation of Robust Zero Energy Extended States}
\author{Jannatul Ferdous$^1$, Cem Yuce$^2$, Andrea Al\`u$^{3,4}$, Hamidreza Ramezani$^1$} 
\email {hamidreza.ramezani@utrgv.edu}
\affiliation{ $^1$Department of Physics and Astronomy, University of Texas Rio Grande Valley, Edinburg, TX 78539, USA \looseness=-4\\
$^2$ Department of Physics, Eskisehir Technical University, Eskisehir, Turkey \looseness=-9\\
$^3$ Photonics Initiative, Advanced Science Research Center, City University of New York, New York, NY, 10031, USA\looseness=-10\\
$^4$ Physics Program, Graduate Center, City University of New York, New York, NY, 10016, USA}

\begin{abstract}
	Topological edge states arise at the interface of two topologically-distinct structures and have two distinct features: they are localized and robust against symmetry protecting disorder. On the other hand, conventional transport in one dimension is associated with extended states, which typically do not have topological robustness. In this paper, using lossy coupled resonators in one dimension, we demonstrate both theoretically and experimentally the existence of robust states residing in the bulk. We show that they are unusually robust against disorders in coupling between adjacent sites and losses. Our work paves the way to a new form of robust transport that is not limited to boundary phenomena and can be accessed more easily from far field.

\end{abstract}

\maketitle
{\it Introduction--} The understanding of topological properties of meta-structures and their consequence in wave matter interaction has enormously expanded over the last decade, see \cite{road} and references in it. %Inspired by the theoretical predictions, there has been remarkable experimental progress in different branches of science such as condensed matter and material physics, electromagnetism [11], electronic circuits , photonics, nonlinear optics, and acoustics [15,21].
One of the most practically relevant aspects of Hermitian topological systems is the existence of eigenstates with unusual protection against disorder. This is a direct consequence of the bulk boundary correspondence, based on which interfaces between bulk materials with different topological invariants necessarily support the presence of confined, highly robust states. In one dimensional systems, this phenomenon corresponds to the presence of a midgap state localized at the interface of two periodic linear arrays, one with nontrivial topology and the other one with trivial topology \cite{joan}. In 2D, a topological edge state allows for edge transport immune to backscattering \cite{Kh, Kim}. In both 1D and 2D, edge states are exponentially localized at the interface. 

The interest in topological states is not limited to Hermitian systems and it has been extended to non-Hermitian lattices \cite{ Henning1,CY, HW, slt, Ulrich,mostafavi,jie4}. This nontrivial extension was inspired by the exotic phenomena, and applications observed in open quantum and classical systems with gain and loss \cite{PT-Rev,zhur}. In particular, non-Hermiticity removes the orthogonality of eigenstates and can support a phase space with exceptional point singularities in which two or more eigenstates and their corresponding eigenvalues become degenerate \cite{Zhang,miri}. Bi-orthogonality and the collapse of Hilbert space under these conditions lead to unconventional dynamical properties, including unidirectional invisibility \cite{Zin}, non-diffracting states \cite{yuce1}, constant intensity waves \cite{mojgan}, flat bands\cite{hamidf}, unconventional lasing and absorption\cite{lg1, h1,h2,h3}, ultra-fast state transfer\cite{sla,fatemeh,jie3}, mode switching\cite{roter, alu22}, coherent perfect absorbers\cite{alu2,jie2}, unidirectional wave manipulation\cite{wen, alu15}, non-reciprocal topological edge state \cite{aluk}, and anomalous phase transition \cite{lg2}, to name a few.

In topological systems, on the other hand, non-Hermiticity makes it possible to go beyond Bloch band theory and observe topologically robust lasers \cite{TL1,TL2}, non-Hermitian skin effect \cite{NHE1, NHE2,NHE3}, robust exceptional points \cite{JY1,JY2}, topologically protected exceptional points \cite{jiet}, complex-energy braiding \cite{Fan,tn}. 

The common theme in both Hermitian and non-Hermitian topological states is localization. The topological eigenstate remains localized when a transition occurs from one topological system to another one with different topology avoiding radiation into the bulk. On the other hand, bulk states are extended all over space and thus are more fragile to disorder. This problem becomes more drastic when unavoidable non-Hermitian disorder  is present, in the form of nonuniformly distributed loss originated from intrinsic material properties or radiation from the structure. Errors in fabrication can affect both intrinsic and radiation losses. While uniform loss can be easily analyzed using an exponential decay term, and in theory extracted via an exponential growing function, disordered non-uniform loss can generate nontrivial scattering responses \cite{RE18, AFT20}. These non-trivial scattering features can be due to resonances solely associated with the non-Hermitian part of the potential and changes in the band diagram of the lattice. Consequently, it can be of great interest to search for nontrivial extended states that are immune to both disorder in Hermitian couplings and non-Hermitian terms in the potential (loss or gain), leading to robust diffraction management.

In this paper, using coupled resonators we propose the realization of a robust extended state corresponding to an extended zero mode originating from the deformation of a topological structure through changing the coupling constants and closing a gap. At first glance, it seems that such a deformation of the lattice causes the bulk boundary correspondence not to be directly applicable to explain the emergence of this robust bulk state. However, we show that due to the deformation in the lattice a zero-mode protected by the symmetry intrudes into the band and consequently becomes de-localized while keeping its robustness. % we show that its presence is due to the deformation the zero-mode protected by symmetry intrudes into the band, which becomes de-localized while keeping its robustness. 
Here, we show theoretically and experimentally that such a robust extended state has resilience against disorder in both Hermitian coupling and non-Hermitian onsite potential. While for our experimental demonstration we use a phononic lattice, a similar phenomenon can be expected to emerge in photonic systems.% specifically in robust large-scale laser arrays with spatially extended laser emission.

{\it Theoretical analysis--} While we will consider in the next section an acoustic waveguide system in a scattering setup to experimentally demonstrate topologically robust bulk states, we start our theoretical analysis with a non-Hermitian variant of Su-Schrieffer-Heeger (SSH) model\cite{ssh} with a nonuniform onsite loss or gain and with open boundary condition. This simple model can predict the physics of our actual experimental setup \cite{Z19,Z20,Z21} and it can be easily adapted to photonic or electronic systems \cite{K17, SSHL}. %The nontrivial nature of the SSH array guarantees the existence of a topological robust edge state with zero energy. 
 The corresponding eigenvalue equation associated with our SSH system, namely ${\cal H}|\psi\rangle={\cal E}|\psi\rangle$, in its spatial representation can be written as
%The Schrodinger-like equation associated with the SSH model and its corresponding Hamiltonian with nonuniform onsite loss or gain can be generally written as 
\begin{equation}
%{\cal E}\psi_j=t_{j-1}(1-\delta_{0,j-1})\psi_{j-1}+t_j(1-\delta_{N,j})\psi_{j+1}+i\gamma_j\psi_j%, \quad j=1,2,3,...,N
{\cal E}\psi_j=t_{j-1}\psi_{j-1}+t_j\psi_{j+1}+i\gamma_j\psi_j, \quad \psi_0=\psi_{N+1}=0
\label{1}
\end{equation}% where $\vec{\Psi}=\left(\psi_1,\psi_2,...,\psi_{N}\right)^T$ is a vector describing the wave function at each site. Here $T$ stand for transpose. Furthermore,
%\begin{equation}
%H_0=\sum_{j=1}^{N-1}t_j(|\psi_j\rangle\langle \psi_{j+1}|+|\psi_{j+1}\rangle\langle \psi_j|)
%\label{2}
%\end{equation}
where $j=1,2,3,...,N$, with $N$ being an odd number, and $t_j=\left\{\begin{array}{cc}
k& \text{if } j \text{ is odd}\\
c& \text{if } j \text{ is even}
\end{array}\right.$. Here, $k$ and $c$ are staggering couplings along the lattice and we assume that $c \leq k $ are both real parameters. %Furthermore, the imaginary part of $\cal H$ is 
%\begin{equation}
%H_I=\sum_{j=1}^{N}\gamma_{j}|\psi_j\rangle\langle \psi_j|
%\label{3}
%\end{equation}
Furthermore, $\gamma_{j}$ indicates the strength of loss ($\gamma_j<0$) or gain ($\gamma_j>0$) in each site. 

The Hermitian SSH model with $\gamma_j=0$, has a bandgap in its spectrum which is equal to $2|k-c|$. Due to the chiral symmetry the corresponding energy eigenvalues come in pairs, namely ${\cal E}=\pm E$. This guarantees that for odd values of $N$, zero energy mode appears in the middle of the band gap. In this case, the eigenmode associated with the zero energy eigenvalue is localized at one edge (here at right side) and its localization length increases as $|k-c|\to 0$ \cite{sc,recentadd}. Thus, the zero mode eigenstate becomes extended when the band gap vanishes at $k=c$. Both localized and extended zero energy modes are robust against coupling disorder since the chiral symmetry remains intact in the presence of disorder in the couplings.

%This means that, for an odd number of sites, which corresponds to a nontrivial geometry, one eigenvalue must appear at zero energy, the so-called zero mode. For an even number of sites corresponding to a nontrivial topological SSH array, a hybridization of two edge states occurs, and only for very long lattices ($N>>0$) two pairs of modes with energies close to zero appear at each edge.
\begin{figure}[h!]
		\includegraphics[width=1.0\linewidth, angle=0]{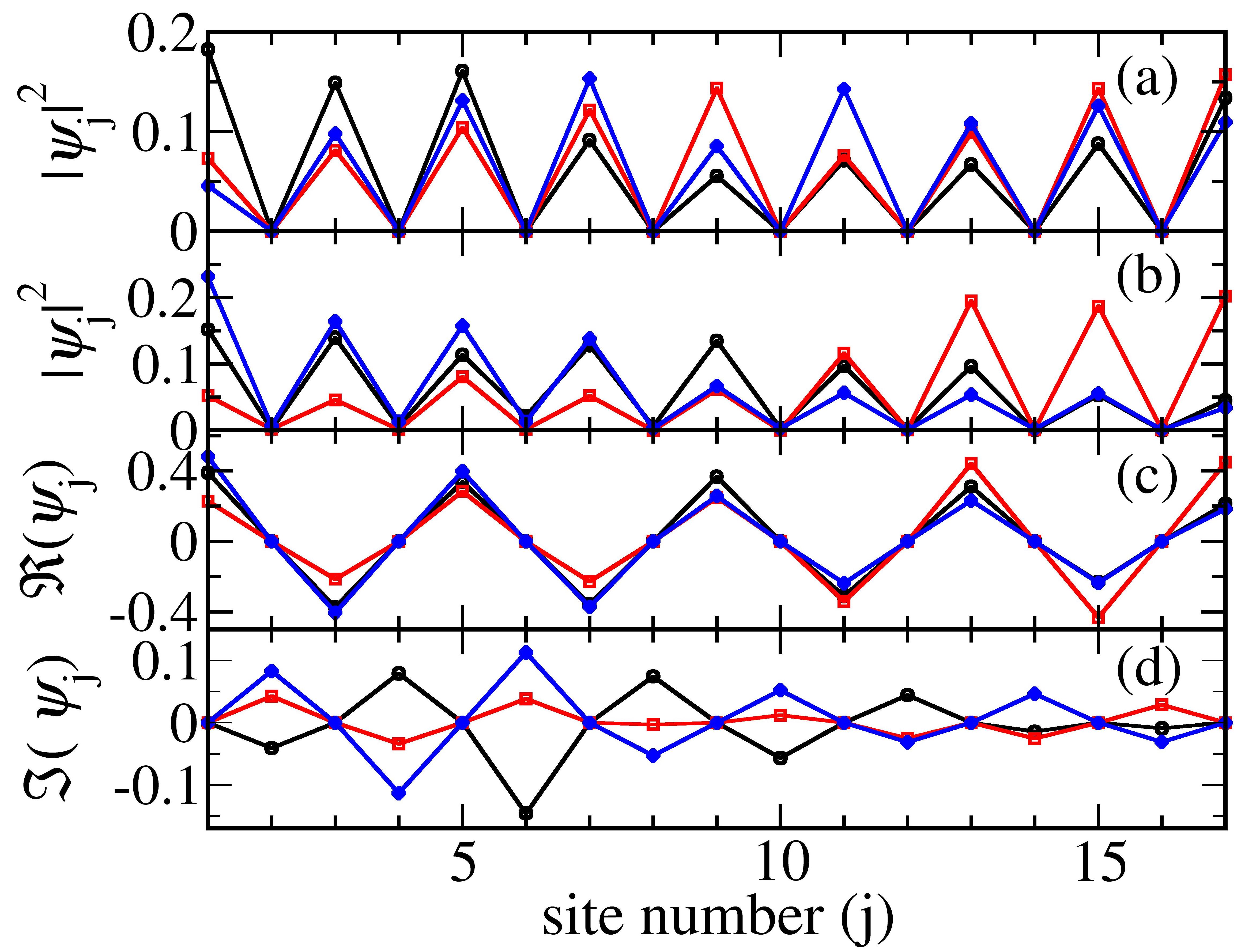}
	\caption{(a) Mode profiles associated with the bulk zero mode in a system described by Hamiltonian $\cal H$ and gain, loss values given by Eq.(\ref{5}) with $20\%$ disorder in the couplings and $\gamma=0.2$. For all realizations the zero mode remains zero in its real part while its imaginary part is equal to $0.2$. (b) The same as (a) but this time $\gamma_{j}$ is selected randomly from a uniform distribution $[ -0.2,0.2]$. We observe that the intensity at even sites is no longer zero. In (c,d) we plot the real and imaginary parts of the curves in (b). We observe that odd sites are purely real and even sites are purely imaginary, confirming the $\pi$ phase shift among odd sites.  }
	\label{fig1}
\end{figure}
%Now let us focus on the zero-mode in a Hermitian SSH system with an odd number of sites. At this time, we consider in particular the scenario of deforming the lattice such that $c\to k$ while we keep track of the zero energy eigenvalue and its corresponding mode. We know that for $c=k$ the energy eigenvalues must come in pairs with the exception of the zero mode. Thus, the zero-energy eigenvalue must remain intact at $c=k$, even in the presence of disorder in the couplings. On the other hand, 
We can find the exact form of the extended zero mode at $k=c$ from Eq.(\ref{1}) 
\begin{equation}
\left\{
\begin{array}{cc}
\psi_{2j-1}=\psi_1& j=1, 3, 5, ...\\
\psi_{2j+1}=-\psi_1& j=1,3,5,...\\
\psi_{2j}=0& j=1,2,3,4,...
\end{array}
\right.
\label{4}
\end{equation}
The zero amplitude in the even sites is a consequence of the $\pi$ phase difference between the nearest neighbors of each even site, which results in a destructive interference at the even sites. From Eq.(\ref{4}) it is clear that the amplitude of the eigenmode associated with the zero mode is equally distributed at the odd sites, while it has zero amplitude on the even sites. Consequently, one can transfer a localized edge state in the Hermitian SSH model to an extended bulk state by deformation of the lattice by means of making $k=c$, while preserving its robustness. %The nonzero amplitude on the 

Now let us discuss the addition of non-Hermitian terms $\gamma_j\neq0$ in Eq.(\ref{1}) when $k=c$. To begin with, let us consider the Hermitian, $H_0$ and non-Hermitian $H_I$ parts of the Hamiltonian associated with our model, $\cal H$. Specifically, $\cal H$ would be an $N\times N$ matrix that its only non-zero elements are ${\cal H}_{j,j+1}={\cal H}_{j+1,j}=k$ and ${\cal H}_{j,j}=i\gamma_j$. 
%\begin{equation*}H_0=\left(\begin{array}{ccccc}
%0 & k& 0 & 0 & ...\\
%k& 0 & k & 0 & ...\\
%0 & k & 0 & k & ...\\
%0 & 0 & k & 0 & ...\\
%... & ... & ... & ... & ...
%\end{array}\right)_{N\times N}
%\end{equation*}
%and 
%\begin{equation*}H_I=i\left(\begin{array}{ccccc}
%\gamma_1 & 0 & 0 & 0 & ...\\
%0 & \gamma_2 & 0 & 0 & ...\\
%0 & 0 & \gamma_3 & 0 & ...\\
%0 & 0 & 0 & \gamma_4 & ...\\
%... & ... & ... & ... & ...
%\end{array}\right)_{N\times N}
%\end{equation*}
%It is easy to show that for 
we can show that for \begin{equation}
\gamma_{j}=(-1)^j\gamma, \quad j=1,2,3,...
\label{5}
\end{equation} we have $\{H_0,H_I\}=0$. The matrix representation of $H_0$The last relation implies that the two Hamiltonians $H_0$ and $H_I$ anti-commute with each other. This anti-commutation condition corresponds to the {\it chiral symmetry} condition for $H_0$. In other words, under constraint (\ref{5}), the non-Hermitian Hamiltonian $H_I$ can be considered as the chiral symmetry operator for the Hamiltonian $H_0$. However, the total Hamiltonian $\cal{H}$ is not chiral symmetric. A consequence of the anti-commutation is that $H_0$ and $H_I$ share a common energy eigenstate. It turns out that this common eigenstate is the zero energy eigenstate of $H_0$ with eigenvalue $-i\gamma$, as it satisfies the aforementioned chiral symmetry. %Notice that the common eigenstate has an eigenvalue different than zero when we apply ${\cal H}=H_0+H_I$ on it. Specifically, for $\cal H$ the real part of the eigenvalue remains zero, while its imaginary part is no longer zero and is equal to $-\gamma$. 
We keep calling this pure imaginary mode as zero mode, as its real part is still zero. This is a unique feature of the zero eigenmode, and there exists no such other eigenstate that is shared between $H_0$ and $H_I$.

In the presence of disorder in the couplings, still the anti-commutation relation is preserved and thus the common zero mode eigenvector remains intact. In this case, the real and imaginary parts of the zero mode associated with $\cal H$ remain intact, and the only quantity that changes is the eigenvector associated with the zero mode. Particularly, we no longer have a uniform eigenvector for the zero mode of $\cal H$. In other words, while we have $\psi_{2j}=0$ as in Eq.(\ref{4}), we will not have $|\psi_{2j+1}|= |\psi_1|$. We have plotted the amplitude at each site associated with the zero mode eigenvalue in Fig.(\ref{fig1}a) for different strengths of disorder in the couplings, which confirms our last statement. Notice that, even for very strong disorder in the coupling we obtain the zero mode, but there is a possibility of having the mode localized and no longer distributed in the bulk. %It is important to note that such localization occurs in the continuum, which is a signature of Anderson localization.

So far it appears that $\psi_{2j}=0$ is the signature of the zero mode independent of the presence of disorder in the couplings and of constraint (\ref{5}) on $H_I$. We introduce now an imaginary part to disorder in the diagonal elements of $H_I$ while we keep the disorder in the coupling elements of $H_0$. By adding such disorder to $H_I$, namely randomly choosing $\gamma_{j}$ from a uniform distribution, the anti-commutation relation between $H_0$ and $H_I$ will not be valid anymore. This is also expected when we consider scattering from a multi-layered medium, since both the imaginary and real parts of the index of refraction play a role in the scattering process. We do not expect that, in the presence of pure imaginary onsite disorder, the zero mode and its corresponding bulk state remain intact. However, our numerical simulations show that the zero mode still has a zero real part, and only its imaginary part is affected by non-Hermitian disorder. To explain this effect, we can use non-degenerate first-order perturbation theory, based on which to first-order approximation the shift in the zero mode eigenvalue is given by 
\begin{equation}
	E^{(1)}_0=i\expval{H_I}{0^{(0)}}
	\label{eqep}
\end{equation}
in which $\ket{0^{(0)}}$ is the eigenvector associated with the zero mode energy of $H_0$ in the presence of disorder in the couplings. Clearly, $E^{(1)}_0$ in Eq.(\ref{eqep}) is purely imaginary, and thus the disorder in the imaginary part of the onsite potentials does not change the real part of the zero mode. This means that the bulk zero mode is robust against disorder in the couplings and its real part is also robust against onsite purely imaginary disorder. 

\begin{figure}
	\includegraphics[width=1.0\linewidth, angle=0]{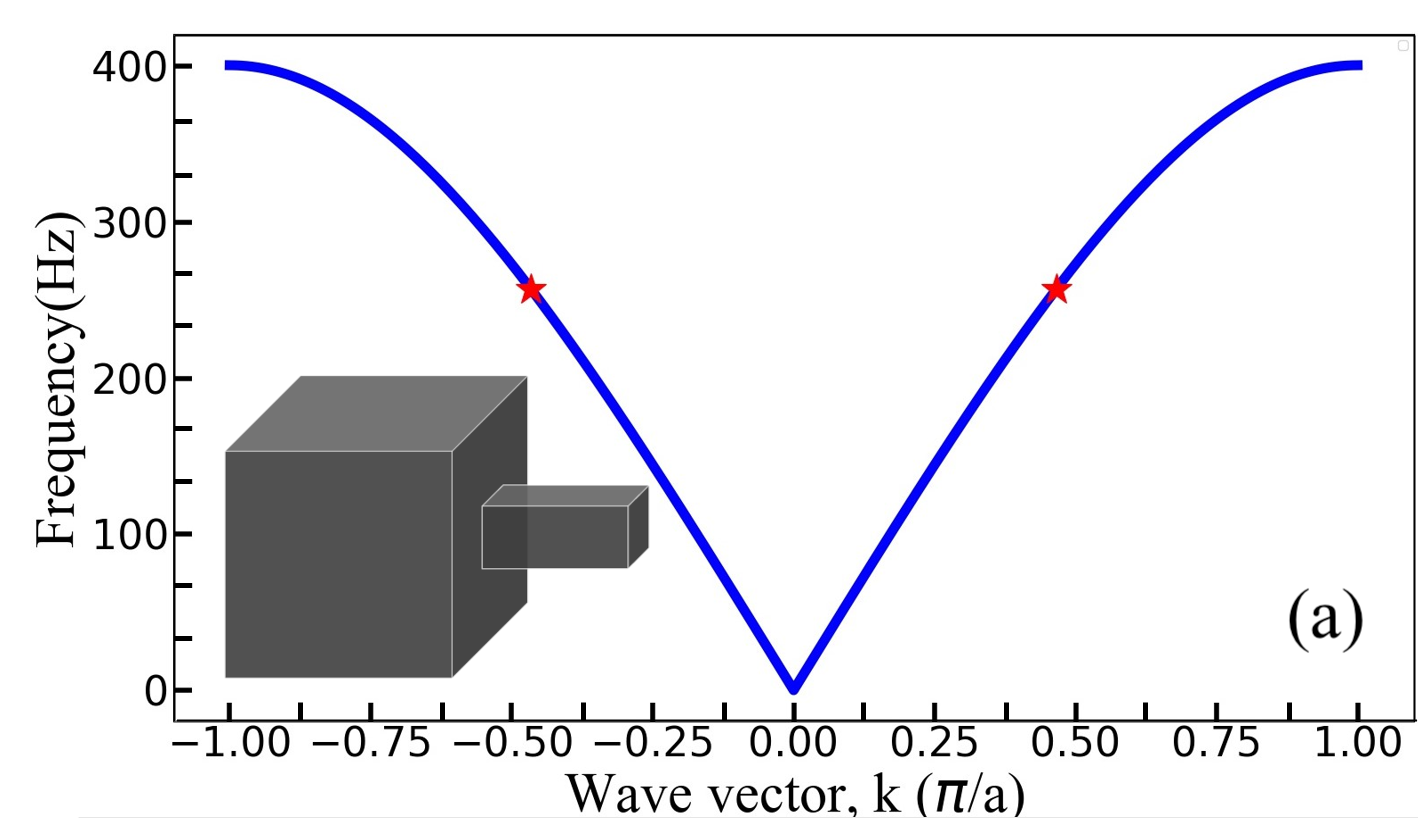}
	\includegraphics[width=1.0\linewidth, angle=0]{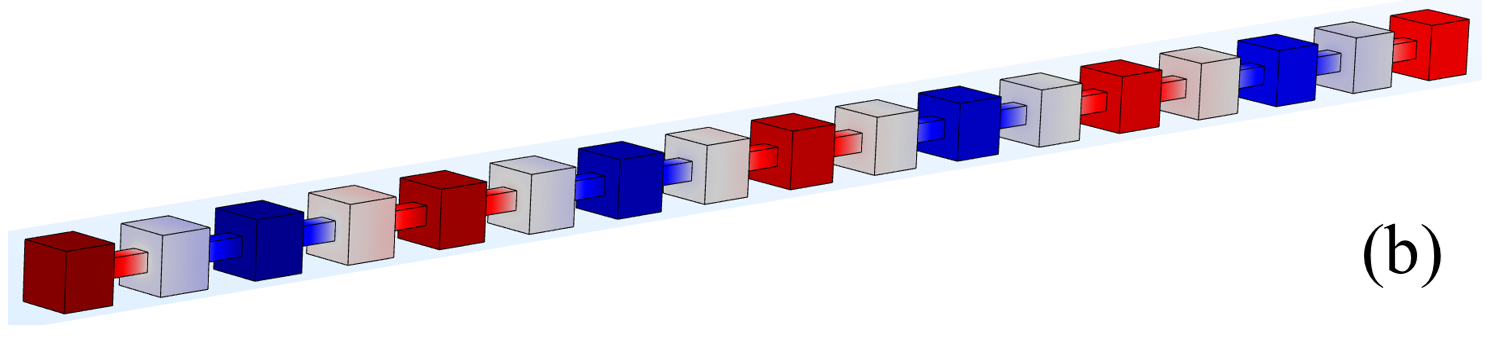}
	\caption{(a) Band-structure for the periodic acoustic waveguide whose unit cell is shown in the inset. The frequency associated with the robust bulk state is identified with the star marks on the band diagram. The unit cell is composed of cuboids with dimension $7 \times 7 \times 7$ cm$^3$ connected to a rectangle with dimension $2.3 \times 2.3 \times 7$ cm$^3$. Our waveguide system is made of such 17 cuboids connected by 16 rectangles. (b) Mode profile of the robust extended mode of the waveguide composed of $17$ cuboids around eigenfrequency $256$ Hz.}
	\label{fig2}
\end{figure}

The above results become more interesting when we consider the eigenvector associated with the zero mode in the presence of purely imaginary onsite potential. 
More specifically, as it is depicted in Fig(\ref{fig1}b), the eigenmode associated with the zero mode does not necessarily have zero intensity in every other site, meaning that $|\psi_{2j}|\neq 0$. Although $|\psi_{2j}|\neq 0$, as it is shown in Fig.(\ref{fig1}c,d) the imaginary part of wave-function at odd sites is zero and thus it is always real while the even sites are always purely imaginary. This means that there is a $\frac{\pi}{2}$ phase difference between nearest-neighboring sites, and thus a $\pi$ phase difference between the sites at odd positions. However, this $\pi$ phase shift does not lead to destructive interference at even sites as now the wave function can be complex at different sites. Thus, we can conclude that the $\pi$ phase shift between the adjacent odd sites is a signature of a robust bulk zero mode even if the mode is distributed all over the system in a non-uniform way.

{\it Experimental observation of robust bulk state--} To demonstrate the predicted robust bulk state experimentally, we use an airborne acoustic waveguide composed of coupled Helmholtz resonators in a scattering setup. As we show later, this waveguide system supports a bulk state with all the properties of the zero mode in the Hamiltonian in Eq.(\ref{1}).

The acoustic system considered here is a two-port hollow waveguide composed of $17$ cuboids resonators with dimension $7\times 7\times 7$ cm$^3$ coupled to each other by rectangles with dimension $2.3\times 2.3\times 2.3$ cm$^3$. The fabricated sample is made of polystyrene foam sheets with density $1050$ $\frac{\text{kg}^3}{\text{m}^3}$ and bulk modulus $16.5$ MPa with an average thickness of $5$ mm. The inner cross-section of the waveguide along the cuboids is kept at $49$ cm$^2$ and along the rectangles is kept at $5.29$ cm$^2$. A similar system has been used previously to generate a tunable non-Hermitian acoustic filter \cite{sr20}.

\begin{figure}
\includegraphics[width=1.0 \linewidth, angle=0]{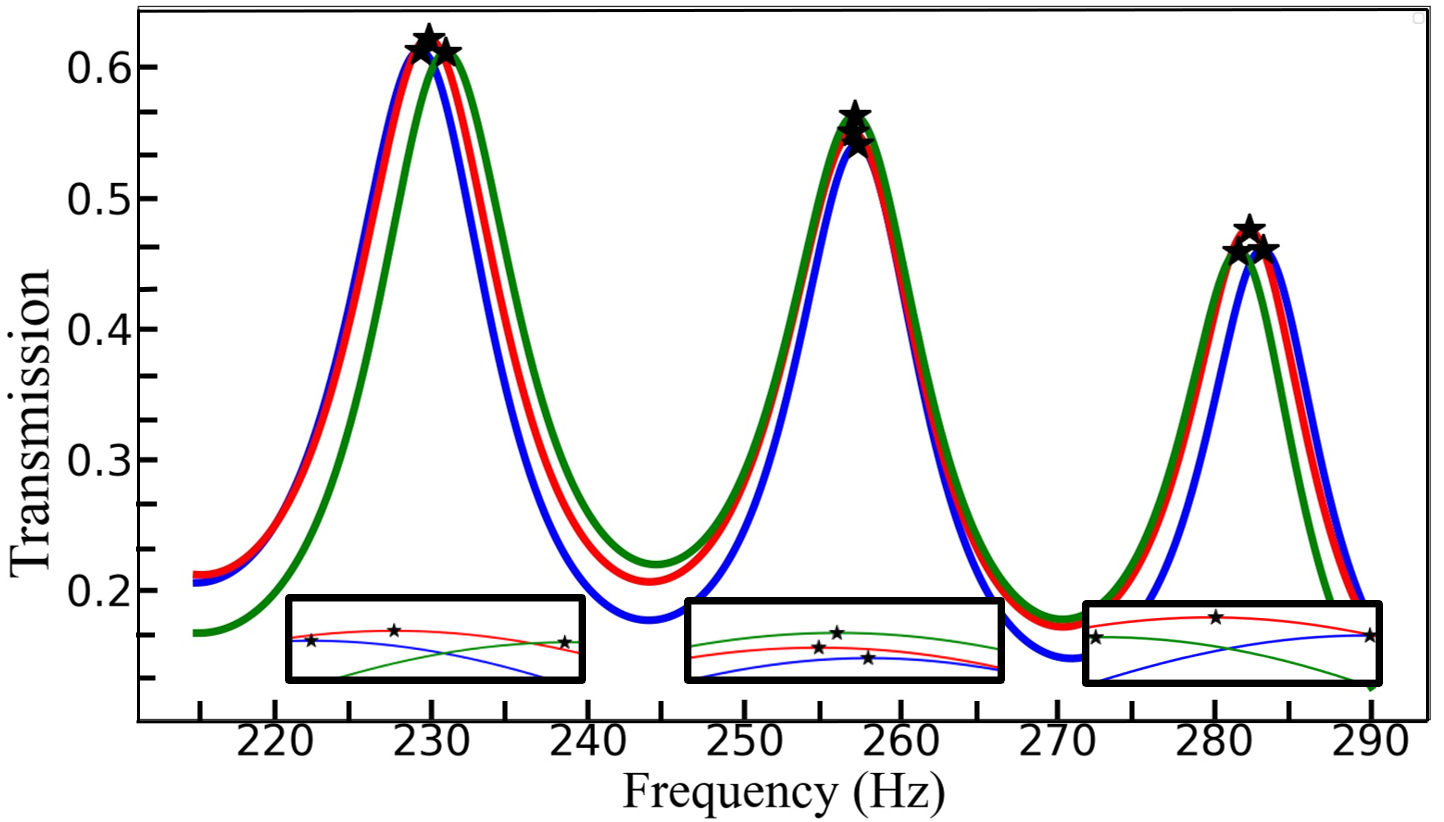}

 	\caption{Full-wave numerical demonstration of topologically protected robust bulk state. % The upper panel shows that the incoming signal is sent from left side and the outgoing signal is received as transmitted signal in the right side of the waveguide. 
 	Blue (red and green) depicts transmission as a function of frequency in the absence (presence) of disorder. Black stars are used to indicate the resonance peaks. As expected the resonance peak associated with the robust mode at frequency $\approx 256$ Hz has negligible shift, while the other peaks have significant shift. The insets zoom in the curves around the peaks.}
 \label{fig3}
\end{figure}

In the search for the robust bulk state, in Fig.(\ref{fig2}a) we plotted the lowest band of this system using numerical full-wave simulations, where we assumed the density of air to be $\rho=1.4$ $\frac{\text {kg}^3}{\text{m}^3}$, while the effective bulk modulus is $\beta=0.931\times10^5 (1+0.01i)$ Pa. The imaginary part of $\beta$ describes the loss in our actual setup. Figure (\ref{fig2}b) confirms that the mode at frequency $256$ Hz, indicated with the stars in Fig.(\ref{fig2}a), has a very similar form as the one described in Eq. (\ref{4}), and thus we expect it to correspond to the robust bulk state at this frequency. To check this claim using full-wave simulation, in Fig. (\ref{fig3}) we plot the transmission associated with the scattering scenario where the system is excited from the left and the signal is collected at the right port. Specifically, the blue curve is the transmission without disorder and uniform loss. The green and red curves correspond to the cases in which we added $10\%$ random disorder to the couplings by changing the length of the cavity sides. We then compared the peaks, indicated with star marks, at different frequencies.  Clearly, as it is also shown in the insets, for the three systems the stars located at the peak associated with the frequency of $256$ Hz are not affected, while the other resonance peaks have a significant shift. 

\begin{figure}
\includegraphics[width=1.0 \linewidth, angle=0]{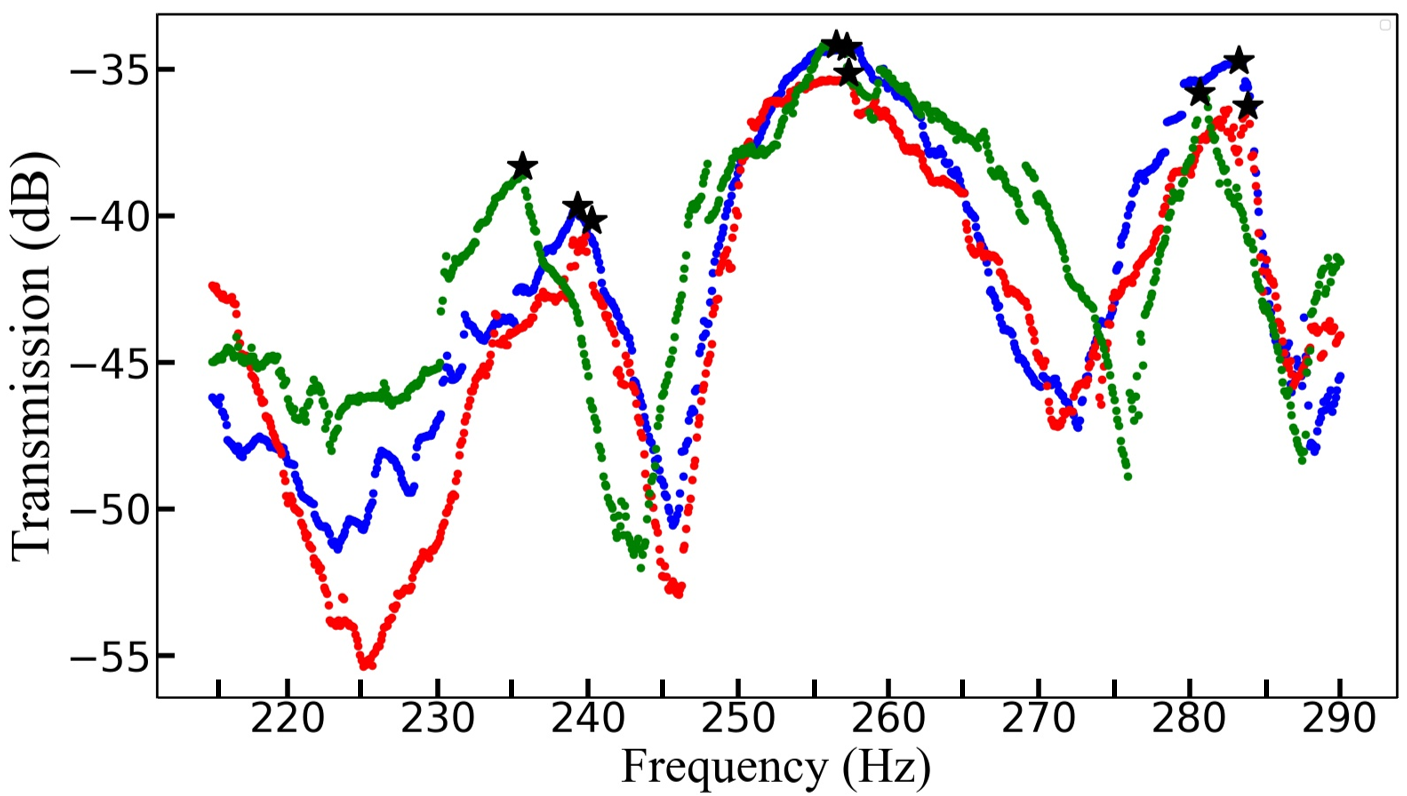}

 	\caption{Experimental verification of robust bulk state. The transmission spectrum for the system without disorder with uncontrolled  small loss is displayed by the blue curve, while the systems with disorder with uncontrolled loss are shown by the red and green curves. Resonance peaks are shown with stars. The robust bulk state is observed around $256$ H,z which has the smallest shift.}
 	\label{fig4}
\end{figure}

The experimental data corresponding to the transmitted signal for a range of frequencies between $215$ Hz and $290$ Hz for three different samples are shown in Fig.(\ref{fig4}). Specifically, the blue dotted curve corresponds to the transmitted signal for the almost perfect system without disorder with almost uniform loss induced by the leakage and absorption of sound through foam sheets along the structure. The red and green dotted curves correspond to the transmitted signal for two different disordered systems where we added $10\%$ disorders to the length of the couplings without moderating the loss. The black stars indicate the resonance peak frequencies in each sample. At the frequency of $256$ Hz, the middle peak of Fig. (\ref{fig4}), less shift in the resonance peak is found in comparison to the other two peaks around $230$ Hz and $280$ Hz, demonstrating the robust feature of the bulk topological state.

To conclude, we have demonstrated the emergence of a protected state against disorder that belongs to the bulk and is not necessarily localized to an edge. The robustness of the bulk state is not affected by coupling disorder and/or disorder in the imaginary part of the potential. While the bulk boundary correspondence cannot directly explain the robustness of the bulk state, the intrusion of the edge state into the band is responsible for the source of the robustness of the bulk state. We believe that the concept of a topologically robust bulk state will offer new perspectives not only in phononics but also in photonics and other application fields. Our results provide a route for developing a new device where direct undisturbed transport of signals is possible.

\begin{acknowledgments}
	H. R. acknowledge the support by the Army Research Office Grant No. W911NF-20-1-0276 and NSF Grant No. PHY-2012172. C. Y.  acknowledge the support from the Scientific and Technological Research Council of Turkey through the 2219 program with grant number 1059B191900044. A. A. has been supported by the Air Force Office of Scientific Research and the Simons Foundation. The views and conclusions contained in this document are those of the authors and should not be interpreted as representing the official policies, either expressed or implied, of the Army Research Office or the U.S. Government. The U.S. Government is authorized to reproduce and distribute reprints for Government purposes notwithstanding any copyright notation herein. 
\end{acknowledgments}

%==================================================================================

\end{document}